\documentclass[twoside,11pt]{article}

\usepackage{jmlr2e}
\usepackage{algpseudocode,multirow, float,bm,amsmath,amsthm,xcolor,colortbl, tikz,pgfplots,makecell, subcaption,tabularx,comment}
\usetikzlibrary{decorations.markings}
\usetikzlibrary{decorations.pathmorphing}
\usetikzlibrary{shapes.geometric, arrows}
\usetikzlibrary{arrows,decorations.pathmorphing,backgrounds,positioning,fit,matrix}
\usetikzlibrary{shapes,decorations,arrows,calc,arrows.meta,fit,positioning}
\usetikzlibrary{pgfplots.groupplots}
\tikzset{
	-Latex,auto,node distance =1 cm and 1 cm,semithick,
	state/.style ={circle, draw, minimum width = 0.7 cm},
	point/.style = {circle, draw, inner sep=0.04cm,fill,node contents={}},
	bidirected/.style={Latex-Latex,dashed},
	el/.style = {inner sep=2pt, align=left, sloped}
}

\newcommand{\md}[1]{\mathbb{#1}}

\newcommand{\thickhline}{%
	\noalign {\ifnum 0=`}\fi \hrule height 1pt
	\futurelet \reserved@a \@xhline
}
\renewcommand{\P}{\mathbb{P}}
\renewcommand{\d}{\mathrm{d}}

\newtheorem{theo}{Theorem}

\newtheorem*{theo*}{Theorem}

\newtheorem{prop}{Proposition}

\DeclareMathOperator\Bern{Bern}

\newcommand\E{\mathbb{E}}

\newcommand\V{\mathbb{V}}

\newcommand\bU{\mathbf{U}}

\newcommand\bx{\mathbf{x}}
\newcommand\bX{\mathbf{X}}

\newcommand\bone{\mathbf{1}}

\usepackage{array}
\newcolumntype{M}[1]{>{\centering\arraybackslash}m{#1}}
\newcolumntype{P}[1]{>{\centering\arraybackslash}p{#1}}
\usepackage{algorithm}

\usepackage{dcolumn}
\newcolumntype{.}{D{.}{.}{-1}}
\newcolumntype{d}[1]{D{.}{.}{#1}}
\newcolumntype{C}{>{$}c<{$}}

\jmlrheading{2}{2023}{1-26}{2/8}{}{Dimitris Bertsimas, Kosuke Imai and Michael Lingzhi Li}

\ShortHeadings{Distributionally Robust Causal Inference}{Bertsimas, Imai and Li}
\firstpageno{1}

\begin{document}

\title{Distributionally Robust Causal Inference with Observational Data}

\author{\name Dimitris Bertsimas \email dbertsim@mit.edu \\
       \addr Sloan School of Management and Operations Research Center\\
       Massachusetts Institute of Technology\\
       Cambridge, MA 02139, USA
       \AND
       \name Kosuke Imai \email imai@harvard.edu \\
       \addr Department of Government and Department of Statistics\\
       Harvard University\\
       Cambridge, MA 02138, USA
        \AND
       \name Michael Lingzhi Li \email mili@hbs.edu \\
       \addr Technology and Operations Management\\
       Harvard Business School\\
       Boston, MA 02163, USA}

\editor{}

\maketitle

\begin{abstract}
  We consider the estimation of average treatment effects in
  observational studies and propose a new framework of robust causal
  inference with unobserved confounders.  Our approach is based on
  distributionally robust optimization and proceeds in two steps.  We
  first specify the maximal degree to which the distribution of
  unobserved potential outcomes may deviate from that of observed
  outcomes.  We then derive sharp bounds on the average treatment
  effects under this assumption.  Our framework encompasses
  the popular marginal sensitivity model as a special case, and we
  demonstrate how the proposed methodology can address a primary
  challenge of the marginal sensitivity model that it produces
  uninformative results when unobserved confounders substantially
  affect treatment and outcome.  Specifically, we develop an
  alternative sensitivity model, called the distributional sensitivity
  model, under the assumption that heterogeneity of treatment
  effect due to unobserved variables is relatively small.  Unlike the
  marginal sensitivity model, the distributional sensitivity model
  allows for potential lack of overlap and often produces
  informative bounds even when unobserved variables substantially
  affect both treatment and outcome.  Finally, we show how to
  extend the distributional sensitivity model to
  difference-in-differences designs and settings with instrumental
  variables.  Through simulation and empirical studies, we demonstrate
  the applicability of the proposed methodology.
\end{abstract}

\begin{keywords}
  average treatment effect, confounding, distributionally robust
  optimization, observational studies, 
  sensitivity analysis
\end{keywords}

\section{Introduction}
\label{sec:Introduction}

Although the unbiased estimates of causal effects can be relatively
easily obtained from experimental data, it is often difficult to
randomize treatment assignment for ethical and logistical reasons.  As
a result, many researchers rely on observational data to ascertain
causal effects.  The primary challenge of such observational studies
is the possible existence of unobserved confounders.  While the
standard approach assumes that the treatment assignment is
unconfounded \citep[e.g.,][]{rosenbaum1983central,robi:rotn:zhao:94},
such an assumption often lacks credibility in real-world applications.

We propose a new methodological framework for causal inference with
observational data under the general observational study settings with
the possibility of unobserved confounding.  Specifically, we formulate
the estimation of average treatment effects in the presence of
unobserved confounders as a distributionally robust optimization
problem under a particular ambiguity set (see, for example,
\cite{robust} and \cite{rahimian2019distributionally} for detailed
reviews of the literature on distributionally robust optimization). We
first specify the maximal degree to which the distribution of
unobserved potential outcomes may differ from that of observed
outcomes.  We then bound the average treatment effects under this
assumption.

The proposed framework includes as a special case of the popular
marginal sensitivity model of \citet{tan2006distributional}, which is
closely related to the sensitivity model of \citet{rose:02c}.  This
marginal sensitivity model has been recently used by several other
researchers
\citep[e.g.,][]{zhao2017sensitivity,kallus2018confounding}.  We show
that the sensitivity analysis based on this model solves the
distributionally robust optimization problem with a particular
ambiguity set.  The construction of this ambiguity set assumes that
the distribution of counterfactual outcome among the treated units is
equal to the weighted distribution of observed outcome among the
control units.

One potential problem with the marginal sensitivity model, however, is
that if unobserved confounders substantially affect the treatment and
outcome, the distribution of potential outcome may not be comparable
between the treated and control units.  Partly for this reason, the
marginal sensitivity model can produce uninformative results in
practice.  To address this issue, we develop an alternative
sensitivity model under our proposed framework, called the
distributional sensitivity model.  The model assumes that the
distribution of counterfactual outcome among the treated units is
similar in its shape (but not in its location) to that of the
distribution of the observed outcome among the same set of units. This
assumption is satisfied if the heterogeneity of treatment effect is
small relative to that of the outcome variable.  A related assumption
is often invoked when estimating heterogeneous treatment effects to
justify the direct modeling of conditional average treatment effect
\citep[e.g.,][]{Hahn2020,Nie2021,Kennedy2022}.

We also show that this distributional sensitivity model can be
extended to other study designs, including the
difference-in-differences, instrumental variables, and settings with
high-dimensional covariates.  Finally, we conduct simulation and
empirical analyses to evaluate the performance of the proposed
methodology.

\noindent {\bf Related literature.}  Much of the previous work on
sensitivity analysis builds upon either the sensitivity model of
\citet{rose:02c} \citep[see
e.g.,][]{tan2006distributional,hasegawa2017sensitivity,
  zhao2017sensitivity,kallus2018confounding, yadlowsky2018bounds,
  foga:20, dorn2021doubly, tan2022model} or regression
models with unobserved variables \citep[see
e.g.,][]{rose:rubi:83a,scharfstein1999adjusting,imbe:03,imai:keel:yama:10,
  cinelli2020making,chernozhukov2022long}.

In contrast, the proposed framework provides an alternative approach
based on the method of distributionally robust optimization.  It
includes the marginal sensitivity model of
\citet{tan2006distributional} as a special case, but can be further
extended to other study designs, including difference-in-differences,
and instrumental variable designs, under the same distributionally
robust optimization framework.  The proposed methodology also exploits
the commonly used assumption that

Another related literature is the one about partial identification
\citep[e.g.,][]{balk:pear:97,mans:07}.  The proposed methodology shows
how a wide range of partial identification results can be obtained
using the distributionally robust optimization approach.  In fact, we
contribute to the growing literature that utilizes robust optimization
for various causal inference problems with partial identification,
mostly in the context of individualized policy learning
\citep[e.g.,][]{benm:etal:21,
  Cui2021_partial,kallus2018confounding,Pu2021,zhan:benm:imai:22}.

\noindent {\bf Organization of the paper.}
The rest of the paper is organized as follows.  In
Section~\ref{sec:low-dimensional}, we introduce the proposed
methodological framework.  We show that this framework incorporates
the marginal sensitivity model as a special case.  We then develop an
alternative sensitivity model, called the distributional sensitivity
model, based on the assumption that the heterogeneity of treatment
effect is small relative to the variation in the outcome across the
treated and control units.  We also extend this framework to the
difference-in-differences as well as the settings with instrumental
variables and high-dimensional covariates.  In
Section~\ref{sec:simulation}, we conduct simulation studies to
evaluate the performance of the proposed methodology.  Finally, in
Section~\ref{sec:empirical} we apply the proposed methodology to three
empirical data sets and compare its performance with some alternative
methods.

\section{Distributionally Robust Causal Inference}
\label{sec:low-dimensional}

In this section, we describe the proposed methodology.  We begin by
presenting our distributionally robust causal inference framework and
then show how it can be applied to various models and settings. 

\subsection{Setup}

Suppose we have a random sample of $n$ units from a target population
$\mathcal{P}$.  We observe the binary treatment $T_i \in \{0, 1\}$, a
$J$-dimensional vector of pre-treatment covariates $\bX_i$, and the
outcome variable of interest, $Y_i \in \mathbb{R}$.  We also have a
vector of unobserved pre-treatment covariates $\bU_i$.  All together,
we assume the tuple $\{\bX_i,\bU_i, T_i, Y_i(1), Y_i(0)\}$ is
independently and identically distributed where $Y_i(t)$ represents
the potential outcome under the treatment assignment $T_i = t$ with
$t \in \{0,1\}$.  The relation between the observed and potential
outcomes is given by $Y_i = T_i Y_i(1)+(1-T_i)Y_i(0)$ where we make
the standard assumption of no interference between units
\citep{rubi:90}.

We consider a general observational study setting, in which both
observed and unobserved pre-treatment covariates, i.e., $\bX$ and
$\bU$, respectively, are potential confounders.  Formally, we assume
that the probability density function (PDF) of the potential outcomes
$Y_i(t)$ and the treatment $T_i$ exist, and can be described by the
following nonparametric models,
\begin{align}
  f_{Y_i(t) \mid \bX_i, \bU_i}(y) \ & = \ \psi_y(t, \bX_i, \bU_i), \\
  \mathbb{P}(T_i = 1 \mid \bX_i, \bU_i) \ & = \ \pi(\bX_i, \bU_i),
\end{align}
for all $y \in \mathbb{R}$.

This model is quite general and makes no functional form assumption
for $\psi$ and $\pi$.  In particular, while all the confounders are
assumed to be included in $\bX_i$ and $\bU_i$, both $\bX_i$ and
$\bU_i$ may also contain variables that only affect either the
treatment or outcome variable, but not both.  We further note that all
results in the paper would apply if we only assume the existence of
the cumulative distribution function rather than the PDF, at the
expense of increased notation to define the quantities of interest in
the paper.

Under this setup, the average treatment effect on the treated subjects (ATT),
which is our quantity of interest throughout this paper, can be
written as,
\begin{equation}
  \tau \ = \ \frac{\int \int  \int y \pi(\bX_i, \bU_i)\{\psi_y(1, \bX_i, \bU_i) - \psi_y(0,
  \bX_i, \bU_i)\} \d y \;\d F(\bU_i) \; \d F(\bX_i)}{\int
  \int\pi(\bX_i, \bU_i) \d F(\bU_i) \; \d F(\bX_i) }. 
\end{equation}

\subsection{The Proposed Framework}

For simplicity, we begin by assuming that the observed pre-treatment
covariates are categorical that take a small number of possible
values.  Under this setting, we can condition on each value of
$\bX_i$, conduct the same analysis, and aggregate the results over the
values of the covariates, allowing us to ignore the existence of the
observed pre-treatment covariates $\bX_i$.  Later, we extend the
proposed methodology to the cases where $\bX_i$ is high-dimensional
and may include continuous covariates (see
Section~\ref{sec:high-dimensional}).

Under this setting, for any given value of the observed pre-treatment
covariates $\bX_i = \bx$, we can define the following four potential
outcome distributions (two distributions defined separately for the
treatment and control groups),
\begin{equation}
  f_{st}(y) \ = \ \P(Y_i(s)=y \mid T_i=t)= \frac{\int \psi_y(s,\bU_i)\{(1-t)+(2t-1)\pi(\bU_i)\} \d F(\bU_i)}{(1-t)+(2t-1)\int \pi(\bU_i) \d F(\bU_i)},
\end{equation}
for $s, t \in \{0, 1\}$ where we simplify the notation by suppressing
the fact that we condition on the observed pre-treatment covariates
$\bX_i = \bx$.

For illustration, we focus on the average treatment effect for the
treated (ATT), which is one of the most common causal quantity of
interests in observational studies.  The ATT can be rewritten as,
\begin{align}
   \tau(p_1,f_{11},f_{01}) 
  \ = \  \left[\int yf_{11}(y) \d y -\int yf_{01}(y) \d y \right] \label{eq:att},
\end{align}
where the general definition of $p_t$ for $t=0,1$ is given by:
\begin{equation*}
	p_t=\P(T_i=t)=(1-t)+(2t-1)\int \pi(\bU_i) \d F(\bU_i).
\end{equation*}
In typical observational studies, the identifiable distributions are
$p_1$, $f_{00}(y)$ and $f_{11}(y)$.  Observational data, however,
provide no information about the distributions of the counterfactual
outcomes, $f_{01}(y)$ and $f_{10}(y)$.  Researchers, therefore, invoke
additional assumptions to point-identify the causal effects of
interest.  For example, the standard assumption of unconfoundedness
implies,
\begin{equation}
    f_{01}(y)=f_{00}(y) \quad \text{and} \quad f_{10}(y)=f_{11}(y) \label{eq:unconfounded}
\end{equation}
for all $y \in \mathbb{R}$
\citep{rosenbaum1983central}. Unfortunately, these assumptions are not
directly testable and hence are often not credible in practice.

Instead, we consider an alternative approach that imposes restrictions
on the relations between the unidentifiable distributions,
$(f_{10}(y), f_{01}(y))$, and the identifiable ones,
$(f_{11}(y),f_{00}(y))$.  There are many such restrictions, and we
show how they can be used to bound the causal effects under the
distributionally robust optimization framework.

Returning to the ATT example, suppose that the degree of
confoundedness due to the unobserved variables $\bU_i$ is assumed to
be sufficiently small.  This assumption can be expressed as the
closeness between the unidentifiable distribution $f_{01}(x)$ and the
identifiable one $f_{11}(y)$ instead of assuming their distributional
equality as done in Equation~\eqref{eq:unconfounded}.  Specifically,
using total variation distance $\delta(\cdot, \cdot)$, we can write
this alternative assumption as,
\begin{equation}
    S^{TV}(\Lambda)\ = \ \{f_{01}(y) \mid
    \delta(f_{01}(y),f_{00}(y))\leq \Lambda\},
\end{equation}
where $\Lambda$ represents the maximal total variation distance between
$f_{01}$ and $f_{00}$.
  
Under the framework of distributionally robust optimization, $S^{TV}$
forms an \emph{ambiguity set} that defines the set of potential
distributions for $f_{01}$. If we wish to obtain the lower bound of
the ATT, we consider the following robust optimization problem:
\begin{equation}
  \widetilde{\tau} \ = \ \min_{f_{01} \in S^{TV}(\Lambda)} \tau(p,f_{11},f_{01}). \label{eq:tau_opt_general}
\end{equation}
The solution, $\widetilde{\tau}$, represents the most conservative ATT
among all the distributions that are consistent with this ambiguity
set.

In general, we may consider any valid ambiguity set $S$ over
$(f_{10},f_{01})$ and define the causal effect of interest as
$\Psi(p_1,f_{11},f_{00},f_{10},f_{01})$.  Suppose, without loss of
generality, that a greater value of the causal effect is desirable.
Then, the distributionally robust causal inference for $\Psi$ under
the ambiguity set $S$ can be written as the following optimization
problem:
\begin{equation}
    \min_{f_{10},f_{01}\in S} \Psi(p_1,f_{11},f_{00},f_{10},f_{01}), \label{eq:dist-robust}
\end{equation}
where the empirical optimization problem replaces
$p_1, f_{11}, f_{00}$ with its empirical estimates.  We next
demonstrate how this proposed distributionally robust causal inference
framework can be applied to various settings.

\subsection{The Marginal Sensitivity Model}
\label{subsec:marginal_sense}

We first show that our distributionally robust causal inference
framework introduced above encompasses the commonly used marginal
sensitivity model of \cite{tan2006distributional} as a special case.
The marginal sensitivity model implies that the odds ratio of
treatment assignment probability with and without conditioning on the
unobserved pre-treatment covariates $\bU_i$ is bounded by the
sensitivity parameter $\Gamma \ge 1$ with probability one,
\begin{equation}
  \frac{1}{\Gamma} \ \le \ \frac{\pi(\bX_i, \bU_i)/(1-\pi(\bX_i,\bU_i))}{e(\bX_i)/(1-e(\bX_i))} \ \le \ \Gamma,
\end{equation}
where $e(\bX_i) = \mathbb{P}(T_i = 1 \mid \bX_i)$.  If we simplify the
notation by suppressing the conditioning on observed confounders
$\bX_i$ as done above, then this equation becomes:
\begin{equation*}
	\frac{1}{\Gamma} \ \le \ \frac{\pi(\bU_i)/(1-\pi(\bU_i))}{p_1/p_0} \ \le \ \Gamma,
\end{equation*}

Let us consider the standard IPW estimator of the ATT, which can be written as:
\begin{equation*}
  \frac{1}{n_1}\sum_{i=1}^n Y_iT_i -\frac{1}{n}\sum_{i=1}^n \frac{\pi(\bU_i)}{1-\pi(\bU_i)} Y_i(1-T_i).
\end{equation*}
where $n_1$ represents the number of treated units and $n$ is the
total number of units.  Under the marginal sensitivity model with the
sensitivity parameter $\Gamma$, robust causal inference based on the
IPW estimator of the ATT can be written as:
\begin{gather}
  \min_{\pi(\bU_i)} \frac{1}{n_1}\sum_{i=1}^n Y_iT_i -\frac{1}{n}\sum_{i=1}^n \frac{\pi(\bU_i)}{1-\pi(\bU_i)} Y_i(1-T_i)\label{eq:inference_msm_orig}, \\
  s.t. \quad \frac{p_1}{\Gamma p_0} \leq \frac{\pi(\bU_i)}{1-\pi(\bU_i)}\leq \frac{\Gamma p_1}{p_0},  \; \; \forall i \label{eq:inference_msm_orig_cons1}\\
  \sum_{i: T_i=0} \frac{\pi(\bU_i)/(1-\pi(\bU_i))}{p_1/p_0} = n_0,\qquad
  \sum_{i: T_i=1} \frac{p_1/p_0}{\pi(\bU_i)/(1-\pi(\bU_i))} = n_1. \label{eq:inference_msm_orig_cons2}
\end{gather}
The following proposition shows that the marginal sensitivity model
can be written as a special case of the proposed distributionally
robust causal inference framework.  All proofs appear in the
supplementary appendix.
\begin{prop}[Marginal sensitivity
  model] \label{prop:marginalsensitivity} The marginal sensitivity
  model given in
  Equations~\eqref{eq:inference_msm_orig}--\eqref{eq:inference_msm_orig_cons2}
  is equivalent to the following distributionally robust causal
  inference problem:
 \begin{equation*}
   \min_{
     f_{01} \in S^{MS}_{01}(\Gamma)} \tau(p,f_{11},f_{01}),
 \end{equation*}
 with the following ambiguity sets,
 \begin{align*}
       &S^{MS}_{01}(\Gamma)=\left\{ f_{01}(y) \mid F_{01}(y)=\widehat{F}_0(y; \mathbf{w}),  \;\;\sum_{i:T_i=0} w_i = 1,\;\;\frac{1}{\Gamma} \leq n_0w_i \leq \Gamma \;\;\forall i \right\},
  \end{align*}
  where $\Gamma \ge 1$ and $\widehat{F}_t(y; \mathbf{w})$ is the
  weighted empirical CDF of observed outcome under the treatment
  condition $T_i = t$ and a vector of observational weights $\mathbf{w}$.
\end{prop}
Proof is given in Appendix~\ref{app:marginalsensitivity}. 

This ambiguity set provides an intuitive interpretation.  For example,
the set of all possible distributions for $f_{01}(y)$ are those
created by weighting samples of $f_{00}(y)$ with an under-sampling or
oversampling factor of at most $\Gamma$. Therefore, the marginal
sensitivity model essentially assumes that the $f_{01}(y)$
distribution is similar to the $f_{00}(y)$ distribution up to some
reweighting.

One important feature of the marginal sensitivity model is that when
modeling the distribution of a counterfactual outcome $f_{01}(y)$, it
uses the information from the distribution of observed outcome under
the control condition, i.e., $f_{00}(y)$, rather than from
$f_{11}(y)$, which is the distribution of observed outcome under the
treatment condition.  This approach will yield a non-informative bound
if the unobserved confounders affect the outcome in a substantial way,
leading to a large value of $\Gamma$.  In such settings, the distance
between $f_{01}(y)$ and $f_{00}(y)$ may be too great for the marginal
sensitivity model to be useful.  We now propose an alternative
sensitivity model, called ``distributional sensitivity model,'' that
addresses this limitation.

\subsection{The Distributional Sensitivity Model}
\label{subsec:syn_confound}

The distributional sensitivity model is based on the assumption that
the distribution of the observed outcome of the treatment group
$f_{11}(y)$ is sufficiently informative about the distribution of its
counterfactual outcome $f_{01}(y)$.  Similarly, if one is interested
in the average treatment effect for the control (ATC), we may assume
that $f_{00}(y)$ is informative about $f_{10}(y)$ for the control
group.  This assumption is credible if the treatment effect is much
smaller than the effect of unobserved confounders on the outcome.  In
such settings, the within-group comparison is likely to be more
informative than the comparison between the treatment and control
groups, which is the basis of the marginal sensitivity model.  In
addition, unlike the marginal sensitivity model, the distributional
sensitivity model allows for the potential lack of overlap between the
treatment and control groups.

Specifically, in the case of the ATT, the distributional sensitivity
model assumes that $f_{01}(y)$ and $f_{11}(y)$ have a similar
distribution up to a location shift $c$.  This type of location shift
model has been used as a justification for the use of rank-sum tests
in randomized experiments \citep{lehm:06}.  The ambiguity set for
$f_{01}(y)$ is given by,
\begin{align}
    S^{DS}_{01}(\Gamma, \delta)& \ = \ \left\{ f_{01}(y) \mid
                                 F_{01}(y)=\widehat{F}_0(y;
                                 \mathbf{w}),  \;\;\sum_{i:T_i=0} w_i
                                 = 1,\;\; 0\leq w_i n_0\leq \Gamma \;\;\forall i,\right.\nonumber\\
    &\qquad\qquad\qquad  \;\;
      \min_c KS(F_{01}(y),F_{11}(y+c)) \leq \delta\Biggl\}, \label{eq:DSambiguity}
\end{align}
where $\Gamma\geq 1$ and $KS(F(y),G(y))=\max_y |F(y)-G(y)|$ represents
the Kolmogorov-Smirnov distance between the distributions $F$ and $G$.
Note that the inequality on $w_i$ along with the normalization of the
weights imply that at least $1/\Gamma$ of the weights would be
non-zero.  Unlike the marginal sensitivity model, we allow the weights
to be zero to account for the potential lack of common support between
$f_{00}(y)$ and $f_{01}(y)$.  We then bound the treatment effect
subject to this ambiguity set.

This ambiguity set for the distributional sensitivity model is similar
to the one used for the marginal sensitivity model.  They both
construct the distribution of counterfactual outcome for the treated
group $f_{01}(y)$ as a weighted distribution of the observed outcome
for the control group $f_{00}(y)$.  The key difference, however, is
that under the distributional sensitivity model, $f_{01}(y)$ is
assumed to have a distributional shape similar to that of $f_{11}(y)$
up to a location shift.

Next, we show that this distributional shape constraint on $F_{01}$
holds if the heterogeneity of treatment effect due to unobserved
variables $\bU_i$ is sufficiently small.  Such an assumption may be
more credible than the assumption of the marginal sensitivity model
that limits the impact of unobserved variables on the treatment and
outcome.

\begin{prop}\label{prop:distributional_bound}
  Assume that $F_{01}$ and $F_{11}$ are minimally $k$-Lipschitz, where
  $k>0$ is the smallest real number that satisfies for all
  $x,y \in \md{R}$:
	\begin{equation*}
		|F_{01}(x) - F_{01}(y)|\leq k|x-y|, \quad
		|F_{11}(x) - F_{11}(y)|\leq k|x-y|
	\end{equation*}
	Then, we have:
	\[\min_c KS(F_{01}(y),F_{11}(y+c)) \leq 3\left(\frac{k\sigma_{01}}{2}\right)^{2/3},\]
	where $\sigma_{01} = \sqrt{\V[Y_i(1)-Y_i(0) \mid T_i=1]}$. 
\end{prop}
Proposition~\ref{prop:distributional_bound} characterizes an upper
bound of the location-shift KS distance in terms of the heterogeneity
of treatment effect due to unobserved variables. In particular, under
mild regularity conditions, the KS distance is small if the outcome
distributions more diffuse than the distribution of treatment effect
(i.e., $k \ll 1/\sigma_{01}$).  Note that in most common distribution
families, $k$ is inversely proportional to the standard deviation.
For example, a minimally $k$-Lipschitz uniform, normal, and
exponential distribution has the variance of exactly $1/12k^2$,
$1/2\pi k^2$, and $1/k^2$, respectively.  Thus, if both $F_{01}$ and
$F_{11}$ follow one of these distributions, the upper bound is given
by,
\begin{equation*}
\min_c KS(F_{01}(y),F_{11}(y+c)) \ \leq \ \begin{cases}
\left(\frac{3\sigma_{01}}{4\sigma}\right)^{2/3} & \text{for uniform} \\
3\left(\frac{\sigma_{01}}{2\sqrt{2\pi}\sigma}\right)^{2/3} & \text{for
normal} \\
3\left(\frac{\sigma_{01}}{2\sigma}\right)^{2/3} & \text{for exponential}
\end{cases},
\end{equation*}
respectively, where
$\sigma=\max\{\sqrt{\V[Y_i(1)\mid T_i=1]}, \sqrt{\V[Y_i(0) \mid
  T_i=1]}\}$.  Thus, in these cases, our KS distance bound scales as
$O\left(\left(\sigma_{01}/\sigma\right)^{2/3}\right)$.  Note
that the $2/3$ exponent can be tightened if we assume the existence of
higher moments of the distribution of the treatment effect.

In sum, our distributional shape constraint holds with a small value
of $\delta$ if the heterogeneity of treatment effect is smaller than
that of the outcome variable, i.e., $\sigma_{01}\ll \sigma$.

Beyond its connection to the heterogeneity of treatment effect,
another distinct feature of the distributional sensitivity model is
the way it uses observed data.  Indeed, at first sight, it may appear
that the distributional sensitivity model utilizes information from
$f_{00}$ and $f_{11}$ in an asymmetric way.  Specifically, we use the
weighted samples from $f_{00}$ to estimate $f_{01}$, while bounding
sampling weights $w_i$ by limiting the difference in shape between
$f_{01}$ and $f_{11}$ after a location shift.  We show, however, that
there exists a duality between bounds on sampling weights and metric
bounds.  This is because a restriction on sampling weights implies a
greater degree of similarity between the two distributions.  The
following theorem establishes that a bound on sampling weights can be
rewritten as a bound on a distance metric between $f_{01}$ and
$f_{00}$.
\begin{theo}[Duality between sampling weights and distance
  metrics]\label{thm:metric_duality} The ambiguity set of the
  distributional sensitivity model, $S^{DS}_{01}(\Gamma, \delta)$ in
  Equation~\eqref{eq:DSambiguity}, can be equivalently characterized
  as:
\begin{align}
    S^{DS}_{01}(\Gamma, \delta)&=\left\{ f_{01}(y) \mid F_{01}(y)=\widehat{F}_0(y; \mathbf{w}), \;\;\sum_{i:T_i=0} w_i = 1, \;\;w_i\geq 0,\right.\nonumber\\
    &\qquad   \qquad d_0(F_{01}(y),F_{00}(y))\leq \frac{\Gamma - 1}{n_0},\ d_1(F_{01}(y),F_{11}(y)) \leq \delta \Biggl\}\label{eq:dual_metric_bound},
\end{align}
where $d_1(F,G)=\min_c KS(F(y),G(y+c))$ is a metric, and
\[d_0(F,G)= \begin{cases}
 \displaystyle \max_y \left|\lim_{\epsilon \to 0^-}
   (F(y)-G(y))-(F(y+\epsilon)- G(y+\epsilon))\right|& 2 \le \Gamma,\\
  \displaystyle \max_y \max\left\{\lim_{\epsilon \to 0^-}
    (F(y)-G(y))-(F(y+\epsilon)- G(y+\epsilon)),0\right\} & 1 \le \Gamma < 2
\end{cases}\] over the space of piecewise constant cumulative
distribution functions, is a quasimetric when $1 \le \Gamma <2$ and a
metric when $\Gamma \geq 2$.
\end{theo}

To implement the distributional sensitivity model, we rewrite
Equation~\eqref{eq:DSambiguity} as a mixed integer linear optimization
problem, which can be solved using existing software packages such as
\cite{gurobi}.  Specifically, for the $KS$ distance metric constraint,
we note that the size of the location shift $c$ is bounded as follows,
\begin{equation*}
  \min_{c \in \mathbb{R}} KS(F_{01}(y), F_{11}(y+c)) \ = \ \min_{c:|c|\leq \max_i Y_i-\min_i Y_i} KS(F_{01}(y),F_{11}(y+c)).
\end{equation*}
Thus, we discretize $c$ with a small value of
$\epsilon:= |\max_i Y_i-\min_i Y_i|/m> 0$, and then optimize over the
set
$L=\{-|\max_i Y_i-\min_i Y_i|,-|\max_i Y_i-\min_i Y_i|+\epsilon,
\cdots, -|\max_i Y_i-\min_i Y_i|+2m\epsilon\}$.  This yields,
\begin{align}
&\min_{c \in \mathbb{R}} KS(F_{01}(y), F_{11}(y+c)) \nonumber \\ \ \approx \ &\min_{j \in \{0,1,\cdots,2m\}} KS(F_{01}(y),F_{11}(y-|\max_i Y_i-\min_i Y_i|+j\epsilon)) \nonumber\\=&\min_{j \in \{0,1,\cdots,2m\}} \max_y |F_{01}(y)-F_{11}(y+c_0+j\epsilon)| \nonumber\\ \ \approx \ & \min_{j \in \{0,1,\cdots,2m\}} \max_{k \in \{0,1,\cdots,2m\}} |F_{01}(\min_i Y_i + k\epsilon )-F_{11}(\min_i Y_i+c_0+(j+k)\epsilon)|, \label{eq:discretize_approx}
\end{align}
where $c_0 =-|\max_i Y_i-\min_i Y_i|$. 
Using this discretization, we can rewrite the optimization problem as
the following mixed-integer linear (feasibility) optimization problem:
\[\min_{\bm{w}, \bm{d}^{ks}, d, \bm{z}}\qquad  \frac{1}{n_1} \sum_{i: T_i =1} Y_i - \sum_{i: T_i =0} w_i Y_i \]
subject to
\begin{align}
	& \sum_{i: T_i=0} w_i =1 \label{eq:ds_eq_full_1}\\
	& 0 \leq w_i \leq \frac{\Gamma}{n_0}  &&\forall i\in\{1,\cdots,n\}  \\
	& d \leq \delta \\
	& d \geq d^{ks}_j -z_j  && \forall j \in \{0,1,\cdots,2m\} \\
	& d^{ks}_j \geq F_{01}(\min_i Y_i + k\epsilon )-F_{11}(\min_i Y_i+c_0+(j+k)\epsilon) && \forall j,k \in \{0,1,\cdots,2m\}\\
	& d^{ks}_j \geq F_{11}(\min_i Y_i+c_0+(j+k)\epsilon) - F_{01}(\min_i Y_i + k\epsilon ) && \forall j, k \in \{0,1,\cdots,2m\}\\
	& \sum_{i} z_j = 2m-1 \quad \text{where} \ z_j \in \{0, 1\} & & \forall j \in \{0,1,\cdots,2m\} \label{eq:ds_eq_full_last}.
\end{align}

\subsection{The Difference-in-Differences Design}
\label{subsec:baseline_outcomes}

We now extend the distributional sensitivity model to the
difference-in-differences (DiD) design where the baseline (i.e.,
pre-treatment) outcome is available for both the treatment and control
groups. We use $Y_b(0,\bU)$ to denote the baseline outcome that is
realized prior to the administration of the treatment.  Under this
setting, therefore, along with $(f_{00}, f_{01}, f_{10}, f_{11})$, we
can define two additional distributions:
\begin{equation}
  f_{bt}(y)=\P(Y_b(0,\bU)=y \mid T(\bU)=t)
\end{equation}
for $t=0,1$.  Note that we observe samples from both distributions
$f_{b0}$ and $f_{b1}$.

Under the DiD design, researchers typically assume the following
parallel trend assumption to estimate the counterfactual outcome mean
for the treated units,
\begin{align}
    \E[Y(0,\bU) \mid T(\bU) = 1]  \ = \ &   \E[Y_b(0,\bU) \mid T(\bU) = 1]
                                   \nonumber \\&+  \{\E[Y(0,\bU) \mid T(\bU) = 0] - \E[Y_b(0,\bU) \mid T(\bU) = 0]\}. \label{eq:did}
\end{align}
This assumption, however, may be violated in practice.  Thus, we
consider the following relaxation by allowing an $\epsilon$ distance
between the distributions of $Y$ on the left and right hand side of
Equation~\eqref{eq:did} under some distance function $d$: 
\begin{equation}
    d(F_{01},F_{b1+00-b0}(y))\leq \epsilon,
\end{equation}
where
$F_{b1+00-b0}(y)=\P(Y_b(0,\bU_1)+Y(0,\bU_0)-Y_b(0,\bU_0) \leq y \mid
T(\bU_1)=1, T(\bU_0)=0)$ is identifiable from the observed data. In
particular, we can choose the function $d$ to be the difference in
expectations, i.e., $d(F(x),G(x)) = |\E_F(x) -\E_G(x)|$, so that it is
interpretable and we can recover the commonly used DiD assumption when
the distance is zero.

This leads to the following distributional sensitivity model,
\begin{equation}
    \min_{f_{01}\in S^{DID}_{01}(\Gamma, \delta,\epsilon )} \Psi(p_1,f_{11},f_{00},f_{10},f_{01}),
\end{equation}
where
\begin{align}
    S^{DID}_{01}(\Gamma, \delta,\epsilon )&=\left\{ f_{01}(y) \mid
                                            F_{01}(y)=\widehat{F}_0(y;
                                            \mathbf{w}),
                                            \;\;\sum_{i:T_i=0} w_i =
                                            1,\right. \nonumber\\
    &\;\; \hspace{.5in}  0\leq w_i \leq \frac{\Gamma}{n_0} \;\;\forall
      i,  \;\; \min_c KS(F_{01}(y),F_{11}(y+c)) \leq \delta,\nonumber \\
                                          &\Biggl.\;\; \hspace{.5in}
                                            d(F_{01}(y),F_{b1+00-b0}(y))\leq \epsilon\Biggr\}.
\end{align}

Furthermore, the above distributional sensitivity model can be
extended to the nonlinear change-in-changes (CIC) model proposed by
\cite{athey2006identification}.  The key assumption of the nonlinear
CIC model can be written as,
\begin{equation}
  F_{01}(y) \ = \ F_{b1}(F_{b0}^{-1}(F_{00}(y))). \label{eq:CICassum}
\end{equation}
We consider a relaxation of this assumption by allowing an $\epsilon$
distance between the distributions on the two sides of
Equation~\eqref{eq:CICassum}. Note that we can estimate
$F_{b1}(F_{b0}^{-1}(F_{00}(y))) $ consistently. Thus, the
distributional sensitivity model for the nonlinear CIC model can be
written as:
\begin{equation}
\min_{f_{01}\in S^{CIC}_{01}(\Gamma, \delta,\epsilon )} \Psi(p_1,f_{11},f_{00},f_{10},f_{01}),
\end{equation}
where
\begin{align}
S^{CIC}_{01}(\Gamma, \delta,\epsilon )&=\left\{ f_{01}(y) \mid
                                        F_{01}(y)=\widehat{F}_0(y;
                                        \mathbf{w}),
                                        \;\;\sum_{i:T_i=0} w_i =
                                        1,\right. \nonumber \\
                                      &\;\; \hspace{.5in} 0\leq w_i \leq \frac{\Gamma}{n_0} \;\;\forall i,
                                        \;\;
                                        \min_c
                                        KS(F_{01}(y),F_{11}(y+c))
                                        \leq
                                        \delta,
 \nonumber \\
                                      &\Biggl.\;\; \hspace{.5in} d(F_{01}(y),F_{b1}(F_{b0}^{-1}(F_{00}(y))\leq \epsilon\Biggr\}.
\end{align}

\subsection{Instrumental Variables}
\label{sec:iv}

We next show that the proposed distributionally robust causal
inference framework can also be applied to instrumental variables
methods \citep{angr:imbe:rubi:96}.  Specifically, we first examine the
estimation of the ATT, which is not identifiable, and then investigate
the robustness of an instrumental variables (i.e., Complier Average
Treatment Effect) estimate to the potential violation of exclusion
restriction.  For simplicity, we consider the settings, in which the
treatment assignment (rather than actual treatment receipt) is either
randomized or unconfounded given a set of pre-treatment covariates.

Specifically, let $Z_i$ represent a binary encouragement variable
which is equal to 1 if unit $i$ is encouraged to receive the treatment
and is equal to 0 otherwise.  We use $T_i \in \{0, 1\}$ to represent
the indicator variable for the actual receipt of treatment.  We use
$T_i(z, \bU_i)$ to denote the potential value of the treatment receipt
variable under the encouragement condition $Z_i = z$ where $\bU_i$
represents the unobserved confounders.  The actual treatment is then
given by $T_i = T_i(Z_i, \bU_i)$.

Throughout this section, along with the unconfoundedness of the
instrument, we assume the following monotonicity assumption, which is
one of the main assumptions of instrumental variables estimation,
\begin{equation}
  T(1, \bU) \ \ge \ T(0, \bU).
\end{equation}
In the current setting, the assumption implies that there is no defier
who would receive the treatment only when they are not encouraged.
Finally, we can define the potential outcome as $Y(t, z, \bU)$ where
the observed outcome is equal to $Y = Y(T, Z, \bU)$. Note that we do
not impose the exclusion restriction assumption, which states that the
instrument affects the outcome only through the treatment.  This
allows us to evaluate robustness against a potential violation of the
exclusion restriction.

Define the following conditional distributions of potential outcomes,
\begin{equation}
  f_{tz}^{t^\prime z^\prime}(y)\ = \ \P(Y(t^\prime, z^\prime,\bU)=y \mid
  T_i(z,\bU)=t, Z =z),
\end{equation}
for $t, t^\prime, z, z^\prime \in \{0, 1\}$.  Note that we can only
identify the distributions $f_{tz}^{t^\prime z^\prime}$ from the
observed data where $t=t^\prime$ and $z=z^\prime$. 

Under this setting, the ATT is defined as,
\begin{equation*}
\text{ATT} \ = \ \frac{p_{11}}{p_1}\left(\int y f_{11}^{11}(y) \d y- \int y f_{11}^{01}(y) \d y\right)  + \frac{p_{10}}{p_1}\left(\int y f_{10}^{10}(y) \d y -\int y f_{10}^{00}(y) \d y\right),
\end{equation*}
where $p_{tz}=n_{tz}/n$, $p_t = p_{t0}+p_{t1}$, and
$n_{tz}=\sum_{i=1}^n \mathbf{1}\{ T_i=t, Z_i=z\}$ denote the sample
size and proportion for each observed strata defined by $T_i=t$ and
$Z_i=z$, respectively, i.e., and.  Thus, we need to estimate the
unidentifiable distributions, i.e., $f_{1z}^{0z}$ with $z=0,1$.

Using our distributionally robust causal inference framework, we use
the identifiable distribution $f_{0z}^{0z}$ to infer the
unidentifiable distribution $f_{1z}^{0z}$ by assuming that
$f_{1z}^{1z}$ has a similar shape as $f_{1z}^{0z}$ for $z=0,1$ up to a
location shift of unknown size. Specifically, the ambiguity set for
$f_{1z}^{0z}$ under the distributional sensitivity model can be given
by:
\begin{align}
  S^{IV}_{z}(\Gamma, \delta)&\ = \ \left\{ f_{1z}^{0z}(y) \mid F_{1z}^{0z}(y)=\widehat{F}_{0z}^{0z}(y; \mathbf{w}),  \;\;\sum_{i:T_i=0, Z_i=z} w_i = 1,\right. \nonumber \\
                          &\hspace{.5in} 0\leq w_i \leq \frac{\Gamma}{n_{0z}} \;\;\forall i: T_i =  0,
                            Z_i=z, \;\; \min_c KS(F_{1z}^{0z}(y),F_{1z}^{1z}(y+c)) \leq \delta \Biggr\},
\end{align}
for $z=0,1$. Under this setup, the standard
exclusion restriction implies the following distributional equality
that couples the unknown distributions,
\begin{align*}
  f^{t^\prime 0}_{tz} = f^{t^\prime 1}_{tz} \quad \text{for} \ t, z, z^\prime \in \{0,1\}.
\end{align*}
This is the distributional representation of the exclusion restriction
assumption that $Z$ only affects $Y$ through $T$. Note that in
particular, the unknown distributions of interest in ATT $f_{1z}^{0z}$
are coupled with the unknown distributions $f_{1z}^{0, 1-z}$ as these
constraints. We relax this assumption by allowing for a small
violation of this assumption.  Specifically, we consider an $\epsilon$
distance between the distributions on the two sides of the equality in
some distance function $d$:
\begin{equation*}
d(F_{1z}^{0z}, F_{1z}^{0, 1-z} )\leq \epsilon \quad \text{for} \ z \in \{0,1\}.
\end{equation*}
Then, robust causal inference under the distributional sensitivity
model that incorporates a relaxed version of exclusion restriction can
be written as,
\begin{align}
& S^{IV}_z(\Gamma, \delta,\epsilon )\nonumber \\ = \ & \Biggl \{  f_{1z}^{0z}(y), f_{1z}^{0, 1-z}(y) \mid F_{1z}^{0z}(y)=\widehat{F}_{0z}^{0z}(y; \mathbf{w}), \;\;  F_{1z}^{0, 1-z}(y)=\widehat{F}_{0, 1-z}^{0, 1-z}(y; \mathbf{w}) \Biggr.\nonumber\\
&\qquad\qquad \sum_{i:T_i=0, Z_i=z} w_i = 1,\;\; \sum_{i:T_i=0, Z_i=1-z} w_i = 1,  \nonumber\\
&\qquad\qquad 0\leq w_i \leq \frac{\Gamma}{n_{01}} \;\;\forall i: T_i=0,Z_i=1,  \;\; 0\leq w_i \leq \frac{\Gamma}{n_{00}} \;\;\forall i: T_i=0,Z_i=0,\\
&\qquad\qquad \min_c KS(F_{1z}^{0z}(y),F_{1z}^{1z}(y+c)) \leq \delta,  \;\; \min_c  KS(F_{1z}^{0, 1-z}(y),F_{1z}^{1z}(y+c)) \leq \delta, \nonumber\\
                                     &\Biggl.\qquad\qquad d(F^{0z}_{1z},F^{0, 1-z}_{1z} )\leq \epsilon\Biggr\}.\nonumber
\end{align}

\subsection{Incorporating High Dimensional Covariates}
\label{sec:high-dimensional}

So far, we have focused on the settings, in which we condition on each
value of covariates $\bX_i$.  In many observational studies, however,
there exist a large number of pre-treatment covariates that need to be
adjusted for.  To handle such cases, we generalize our
distributionally robust causal inference framework by directly
controlling the differences in the observed covariates between the
treatment and control groups.

Specifically, suppose that we have a total of $J$ covariates.  Let
$X_{ij}$ denote the $j$th covariate for unit $i$ where
$j=1,2,\ldots,J$, and $g_{t}(X_{ij})$ represent the density function
of $X_{ij}$ for the units with the treatment status $T_i = t$ where
$t=0,1$.  We use $C(x_{j})$ to denote the ambiguity set for the $j$th
covariate.  Then, the general optimization problem under our
distributionally robust causal inference framework becomes,
\begin{equation}
  \min_{\substack{f_{10}(y),f_{01}(y)\in S(y) \\  g_{0}(x_j),g_{1}(x_j)\in C(x_j) \; \forall j}} \Psi(p_1,f_{11},f_{00},f_{10},f_{01}),
\end{equation}
where $S(y)$ is the distributionally robust set for the unobserved
counterfactual outcomes (as in
Sections~\ref{subsec:marginal_sense}~and~\ref{subsec:syn_confound}),
and $C(x_j)$ are the ambiguity sets in which we
limit the difference in the distribution of $x_j$ under $T=0$ and
$T=1$ to ensure covariate matching. 

For illustration, we revisit the distributional sensitivity model for
the ATT considered in Section~\ref{subsec:syn_confound}.  Under the
high-dimensional setting, we may require $C(x_j)$ to be such that the
$L_1$ distance between first moments of $\bX$ between $f_{11}$ and
$f_{01}$ to be less than $\epsilon$ \citep{zubizarreta2012using}:
\begin{equation}
    \sum_{j=1}^J \left |\frac{1}{n_1}\sum_{i: T_i=1}
      X_{ij}-\frac{1}{n_0} \sum_{i: T_i=0}w_iX_{ij}\right |\leq \epsilon. \label{eq:first_moments_cons}
\end{equation}
The equivalent Lagrangian form is,
\begin{equation}
    \min_{\bm{w} \in S^{DS}_{01}(\Gamma,\delta)}  \frac{1}{n_1}
    \sum_{i: T_i=1} Y_i- \frac{1}{n_0} \sum_{i:T_i=0} w_iY_i + \lambda
    \sum_{l=1}^J \left |\frac{1}{n_1}\sum_{i: T_i=1}
      X_{ij}-\frac{1}{n_0} \sum_{i: T_i=0}w_iX_{ij}\right |. \label{eq:first_moments_obj}
  \end{equation}
  We can then utilize this Lagrangian objective function and implement
  the constraints for the distributional sensitivity model as detailed
  in Equations~\eqref{eq:ds_eq_full_1}--\eqref{eq:ds_eq_full_last}.

\section{Simulation Experiments}
\label{sec:simulation}

We conduct a simulation study to understand the empirical performance
of the distributional sensitivity model. Specifically, we generate the
data based on the following potential outcomes model,
\begin{align*}
    T_i(u) &= \Bern(0.6u + 0.2),\\
    Y_i(t, u) &= (1-u)(t-0.5)\nu_i + u(t-0.5)\eta_i+ \theta_i + \epsilon_i,
\end{align*}
where $v \sim N(0,1)$, $\nu_i \sim N(\tau_1,1)$,
$\eta_i \sim N(\tau_2,1)$, $u \sim \Bern(p)$, $\theta_i \sim N(0,2)$
and $\epsilon_i \sim N(0,0.1)$. Here, $u$ is an unobserved confounder
such that when $u=1$, $\P(T=1)=0.8$ and
$\E[Y(1)-Y(0) \mid u=1]=0.8\tau_2+0.2\tau_1$, and when $u=0$,
$\P(T=1)=0.2$ and $\E[Y(1)-Y(0) \mid u=1]=0.2\tau_2+0.8\tau_1$. When
$\tau_2>\tau_1$, $u$ assigns a higher treatment probability to an
individual with a higher treatment effect, and vice versa when
$\tau_2<\tau_1$. We also note that by design, all of the
$f_{11},f_{00},f_{10}, f_{01}$ distributions are different so that no
framework has a clear advantage.

The causal quantity of interest is the average treatment effect on the
treated (ATT), which can be written as the following function of model
parameters:
\begin{equation*}
	\E[Y(1)-Y(0) \mid T=1]=\frac{8p\tau_2+2(1-p)\tau_1}{6p+2}.
\end{equation*}
We create multiple scenarios based on different values of $(\tau_1,\tau_2,p)$:
\begin{enumerate}
	\item $(\tau_1,\tau_2,p) = (2,3,0.5)$,
	\item $(\tau_1,\tau_2,p) = (3,2,0.5)$,
	\item $(\tau_1,\tau_2,p) = (2,3,0.8)$.
\end{enumerate}
For each scenario, we generate a  dataset with
$n \in \{100, 200, 500\}$ samples over 1,000 independent runs. For
each run, we examine the distributional sensitivity model of
Section~\ref{subsec:syn_confound} with $\delta = 0.1$ and the marginal
sensitivity model across different values of $\Gamma$ (we include
experiments on other $\delta$ values in
Appendix~\ref{app:synthetic_delta}).  Across both sensitivity models,
the value of $\Gamma$ is comparable because it provides identical
upper bounds on the weights $w_i$. 

\begin{table}[t!]
	\centering\setlength{\tabcolsep}{2.5pt}
	\resizebox{\textwidth}{!}{
		\begin{tabular}{l|..|..|..}
			\hline
			&  \multicolumn{2}{c|}{$\bm{n=100}$} &  \multicolumn{2}{c|}{$\bm{n=200}$}& \multicolumn{2}{c}{$\bm{n=500}$}\\\hline
			& \multicolumn{1}{c}{bias} &
			\multicolumn{1}{c|}{s.d.}
			& \multicolumn{1}{c}{bias} &
			\multicolumn{1}{c|}{s.d.}
			& \multicolumn{1}{c}{bias} &
			\multicolumn{1}{c}{s.d.}  \\ \hline
			\multicolumn{1}{l|}{\textbf{Scenario 1}}  & & & & & &\\
			\multicolumn{1}{l|}{\textit{Distributional Sensitivity}}  & & & &&&\\
			\quad $\Gamma = 2$  &-1.551 & 0.540  & -1.704 & 0.369   &  -1.795  & 0.231  \\
			\quad $\Gamma = 3$ & -1.889  & 0.638  & -2.124 & 0.422    &  -2.267&  0.258 \\
			\quad $\Gamma = 5$   & -2.255 & 0.727  & -2.576 & 0.487  &  -2.741  &  0.309   \\
			\multicolumn{1}{l|}{\textit{Marginal Sensitivity}}  & & & &&&\\
			\quad $\Gamma = 2$  & -1.938 & 0.465  & -1.954 & 0.316  &  -1.944  & 0.207   \\
			\quad $\Gamma = 3$ & -2.506  & 0.470  & -2.506 & 0.348    &  -2.550 &  0.218   \\
			\quad $\Gamma = 5$   & -3.164 & 0.529  & -3.175 & 0.391  &  -3.192  &  0.241   \\\hline
			\multicolumn{1}{l|}{\textbf{Scenario 2}}  & & & & & &\\
			\multicolumn{1}{l|}{\textit{Distributional Sensitivity}}  & & & &&&\\
			\quad $\Gamma = 2$  & -0.930 & 0.539  & -1.079 & 0.381   &  -1.187 & 0.228  \\
			\quad $\Gamma = 3$ & -1.343  & 0.659  & -1.545 & 0.433    &  -1.664 &  0.261   \\
			\quad $\Gamma = 5$   & -1.659  & 0.727  & -1.982 & 0.521 &  -2.137  &  0.315   \\
			\multicolumn{1}{l|}{\textit{Marginal Sensitivity}}  & & & &&&\\
			\quad $\Gamma = 2$  & -1.321 & 0.455  & -1.330 & 0.321  &  -1.333  & 0.208  \\
			\quad $\Gamma = 3$ & -1.916  & 0.474  & -1.953 & 0.339    & -1.973 &  0.216   \\
			\quad $\Gamma = 5$   & -2.518 & 0.546  & -2.577 & 0.394  &  -2.594  &  0.233   \\\hline
			\multicolumn{1}{l|}{\textbf{Scenario 3}}  & & & & & &\\
			\multicolumn{1}{l|}{\textit{Distributional Sensitivity}}  & & & &&&\\
			\quad $\Gamma = 2$  & -1.362 & 0.672  & -1.569 & 0.430  &  -1.677  & 0.246   \\
			\quad $\Gamma = 3$ & -1.671  & 0.747  & -1.975 & 0.496    & -2.150 &  0.305   \\
			\quad $\Gamma = 5$   & -1.891  & 0.785  & -2.360 & 0.590 & -2.631  &  0.371   \\
			\multicolumn{1}{l|}{\textit{Marginal Sensitivity}}  & & & &&&\\
			\quad $\Gamma = 2$  & -1.847 & 0.513  & -1.870 & 0.345  &  -1.865  & 0.218   \\
			\quad $\Gamma = 3$ & -2.466 & 0.549  & -2.455 & 0.392    & -2.475 &  0.244   \\
			\quad $\Gamma = 5$   & -3.047 & 0.598  & -3.090 & 0.435  &  -3.092  &  0.271   \\\hline\hline
		\end{tabular}
	}
	\caption{A Simulation Study of Distributional Sensitivity
          (with $\delta = 0.1$) and Marginal Sensitivity Models.  The
          table presents the estimated bias and standard deviation of
          the ATT estimate for both sensitivity models under
          difference scenarios, robustness ($\Gamma$) and number of
          samples $n$.} \label{tb:synthetic_ms_compare}
\end{table}

Table~\ref{tb:synthetic_ms_compare} presents the bias and standard
deviation of the ATT estimate for distributional and marginal
sensitivity models, using their respective lower bound.  As expected,
both methods return conservative estimates of the ATT. We also find
that the bias remain relatively similar across different
scenarios. However, the distributional sensitivity models generally
produce less conservative results compared to the marginal sensitivity
models, with a growing difference as we increase $\Gamma$.
Appendix~\ref{app:synthetic_delta} shows that this result still holds
even if we increase $\delta$. In particular, at high levels of
robustness, the distributional sensitivity model is able to estimate
the ATT with a bias whose magnitude is half of the corresponding bias
for the marginal sensitivity model.  This shows that by using both
$f_{00}$ and $f_{11}$, the distributional sensitivity model is able to
generate less conservative estimates under a wide range of parameter
values.

\section{Empirical Evaluations}
\label{sec:empirical}

For empirical demonstration, we apply the proposed distributionally
robust causal inference framework to the following diverse data sets:
\begin{enumerate}
\item \textbf{National Supported Work Demonstration (NSW) Dataset}:
  This is a well-known data set, which was originally used by
  \cite{lalonde1986evaluating} to evaluate the accuracy of various
  causal inference methods for observational studies.  The data set
  has an experimental benchmark which we use to assess the performance
  of our methodology.
      
\item \textbf{Boston Medical Center (BMC) Diabetes Dataset}: We apply
  the distributional sensitivity model to this observational study and
  demonstrate how our model avoids a potentially erroneous medical
  conclusion.
      
\item \textbf{Student/Teacher Achievement Ratio (STAR) Dataset}: Like
  the analysis of the NSW data, we use the setup from
  \cite{wilde2007close} to construct a synthetic observational dataset
  from the original randomized STAR dataset.  We then evaluate the
  performance of our methodology using the experimental estimate as a
  benchmark.

\end{enumerate}

\subsection{NSW Dataset}
\label{subsec:nsw}

The National Supported Work Demonstration (NSW) was a temporary
employment program to help disadvantaged workers by giving them work
experience and counseling in a sheltered environment. Specifically,
the NSW randomly assigned qualified applicants to treatment and
control groups, where workers in the treatment group were given a
guaranteed job for 9 to 18 months. The primary outcome of interest $Y$
is the (annualized) earnings in 1978, 36 months after the program. The
experimental data in total contains $n=722$ observations, with
$n_1=297$ participants assigned to the treatment group and $n_0=425$
participants in the control group. There are 7 available pre-treatment
covariates $\bm{X}$ that records the demographics and pre-treatment
earnings of the participants.

We consider the setup of \cite{lalonde1986evaluating}, where
non-experimental control groups were constructed from the Population
Survey of Income Dynamics (PSID) and Current Population Survey
(CPS). We replace the experimental controls with the non-experimental
controls and apply the distributional sensitivity model
$S^{DS}_{01}(\Gamma, \delta)$ to robustly infer the ATT.  Since the
treated units come from the randomized experiment, we use the
estimated average treatment effect of $\$886$ in the original
experiment as a benchmark ATT estimate.

To prevent numeric issues due to the wide range of incomes, we
log-transform the outcome variable before applying our methodology. We
also impose matching on the first moments using
Equation~\eqref{eq:first_moments_obj} with $\lambda=1000$, which was selected so that the characteristic magnitude of the Lagrangian objective is significantly larger than the ATT objective, enforcing best-possible first moment matching. We chose $\Gamma$ such
that the minimum number of selected
samples is 30, 60, and 100, corresponding to a $\Gamma$ of 25, 12, and
8 respectively. We further vary $\delta$ between $0.02$ and $0.05$ as
without any weighting, the distributional KL distance between $Y(1)$
and $Y(0)$ is just over 0.2.  Lastly, we consider the distributional
sensitivity model under the difference-in-differences design as
described in Section~\ref{subsec:baseline_outcomes}, where we treat
the 1975 income as the baseline outcome, and utilize the mean
difference metric $d(F,G)=|\mu(F)-\mu(G)|$.

\begin{table}[t]
  \centering
  \begin{tabular}{|c|ccc|...|}
    \hline \multicolumn{1}{|c|}{\textbf{Model}} & \multicolumn{1}{c}{$\bm{\Gamma}$} & \multicolumn{1}{c}{$\bm{\delta}$} & \multicolumn{1}{c|}{$\bm{\epsilon}$} & \multicolumn{1}{c}{\textbf{Estimate}} & \multicolumn{1}{c}{\textbf{s.e.}} & \multicolumn{1}{c|}{\textbf{Bias}} \\\hline
    \multirow{5}{*}{Distributional sensitivity model} & 25 & 0.02 & &  -54.9 & 1585.0 & -940.9\\
                           & 12 & 0.02 &  & -163.0 & 1320.4 & -1049.0\\
                           &8 & 0.02 & & -294.5 & 1166.2 & -1180.5\\
                           &25 & 0.03 & & -304.3 & 1586.7 & -1190.3\\
                           & 25 & 0.05 &  & -559.2 & 1592.8 & -1445.2\\\hline
    \multirow{5}{*}{\makecell{Distributional sensitivity model \\under the
    difference-in-differences}} & 25 & 0.02 & 100 & 320.7 & 1585.7 & -565.3\\
                           & 25 & 0.02 & 200& 220.7 & 1585.6 & -665.3\\
                           &25 & 0.02 & 500& -54.9 & 1585.0 & -940.9\\
                           &25 & 0.03 &500 & -79.2 & 1587.5 & -965.2\\
                           & 25 & 0.05 &500 & -79.3 & 1591.0 & -965.3\\\hline
    Regression \citep{lalonde1986evaluating} & & & &-1228.0 & 896.0 & -2114.0\\ \hline
  \end{tabular}
  \caption{Results of robust inference under the distributional
    sensitivity model for the NSW dataset with PSID controls. The
    table presents the estimate, standard error, and bias of the
    robust estimator under different robustness parameters $\Gamma$
    and $\delta$. } 
  \label{tab:nsw_results_psid}
\end{table}

The results for the PSID dataset are shown in
Table~\ref{tab:nsw_results_psid}. For the standard errors under the
distributional sensitivity model, we calculate the standard errors by
treating the weights as fixed and computing the conditional weighted standard error.  It is
well known that the experimental results based on the PSID datasets
are difficult to recover \citep{dehejia1999causal}. We observe that
linear regression model produces large negative biases relative to the
benchmark value. In contrast, the distributional sensitivity model is
designed to be conservative, but generally produces smaller biases
across a wide range of robustness parameters.  In particular, under
the difference-in-differences design, the distributional sensitivity
model yields even smaller bias.
  
\begin{table}[t]
    \centering
    \begin{tabular}{|c|...|...|}
    	\hline \multirow{2}{*}{\textbf{Dataset}} &
                                                   \multicolumn{3}{c|}{\textbf{Distributional sensitivity model}} & \multicolumn{3}{c|}{\textbf{Linear regression}} \\\cline{2-7}
          &  \multicolumn{1}{|c}{\textbf{Estimate}} & \multicolumn{1}{c}{\textbf{s.e.}} & \multicolumn{1}{c|}{\textbf{Bias}}   &  \multicolumn{1}{c}{\textbf{Estimate}} & \multicolumn{1}{c}{\textbf{s.e.}} & \multicolumn{1}{c|}{\textbf{Bias}}\\\hline
         PSID  & -54.9 & 1585.0 & -940.9  & -1228.0 & 896.0 & -2114.0\\
         CPS & -252.6 & 1003.7 & -1138.6 & -805.0 & 484.0 & -1691.0\\
         CPS2 & -190.1 & 933.6 & -1076.1 & -319.0 & 761.0 & -1205.0\\
         CPS3 & 408.2 & 922.4 & -477.8 & 1466.0 & 984.0 & 580.0\\\hline
    \end{tabular}
    \caption{Results of robust inference under the distributional
      sensitivity model and linear regression for the NSW dataset with different control
      data sets from \cite{lalonde1986evaluating}. The linear regression results are from \cite{lalonde1986evaluating}. We choose
      $\Gamma=n_1/100$ and $\delta=0.02$ for these experiments.  The
      table presents the estimate, standard error, and bias of the
      robust estimator. }
    \label{tab:nsw_results_all}
\end{table}

To further illustrate the robustness of our results, we apply the
distributional sensitivity model to the other non-experimental
datasets from
\cite{lalonde1986evaluating} (PSID2 and PSID3 were removed due to data issues), and compare our results with the resulted from the linear regression controlled on all variables from \cite{lalonde1986evaluating}. Table~\ref{tab:nsw_results_all} shows
that our estimates are again conservative, as expected, but the biases
are all approximately smaller than the standard error. The absolute magnitude of the biases from our methodology is consistently lower than the bias achieved through regression.  Together with the results in
Table~\ref{tab:nsw_results_psid}, our findings demonstrate that the
distributional sensitivity model is able to deliver robustness
guarantees without incurring a large bias.

\subsection{BMC Dataset}
\label{subsec:bmc}

We next apply the distributional sensitivity model to the BMC
dataset. The BMC dataset consists of visit records of 10,806 Type II
diabetic patients from 1999 to 2014. Specifically, during each visit,
the doctor would prescribe a treatment regimen for the patient for the
next treatment period. The outcome variable is the change in
Hemoglobin A1c (HbA1c), a key indicator of blood glucose, between the
visit and the end of the treatment period.  A negative outcome value
is desirable because the key objective of the treatment regimes is to
lower the blood glucose levels.

To avoid the complications due to potential carryover effects between
visits, we focus only on the set of first patient visits.  We
construct a binary treatment by focusing on two
commonly-prescribed treatments: the Metformin monotherapy (MET) and
the Insulin monotherapy (INS). In total, we have $n=2,538$ valid
visits, where $n_1=1869$ visits prescribed Metformin and $n_0=669$
prescribed insulin. There are 22 pre-visit covariates that describe
demographics, treatment history, and comorbidities of the patient.

We are interested in the ATT as well as the average treatment effect
for the control (ATC). We apply the distributional sensitivity model
while balancing the first moments of all covariates with
$\lambda = 1000$ to enforce best-possible first moment matching. We
choose the same sensitivity parameter values, $\Gamma=n_1/100$ and
$\delta=0.02$, as those used for the NSW study. This ensures 100
minimum samples are selected and the maximum distributional KL
distance is feasible. Lastly, we compare our results to those of
Causal Forests \citep{wager2018estimation} and of the classic
doubly-robust estimator from \cite{robins1994estimation} where we
utilize Gradient Boosted Trees for both the propensity and the outcome
models.

\begin{table}[t]
  \centering
  \resizebox{\textwidth}{!}{
    \begin{tabular}{|p{2.8cm}|ccc|ccc|}
      \hline \multicolumn{1}{|c|}{\textbf{Method}} & $\bm{\E[Y(1) \mid T=1]}$ & $\bm{\E[Y(0) \mid T=1]}$ & \textbf{ATT} &  $\bm{\E[Y(1) \mid T=0]}$ & $\bm{\E[Y(0) \mid T=0]}$ & \textbf{ATC}  \\\hline
      Doubly Robust &  $-0.87$ & $-0.12$ & $-0.75$ & $-1.66$ & $-0.89$ & $-0.77$\\
      Causal Forests &  $-0.87$ & $-0.17$ & $-0.70$ & $-1.84$ & $-0.89$ & $-0.96$\\\hline
      \makecell[l]{Distributional \\ sensitivity model} &  $-0.87$ & $-0.51$ & $-0.36$ & $-0.72$ & $-0.89$ & $0.17$\\\hline 
    \end{tabular}
  }
  \caption{Results of the distributional sensitivity model on the BMC
    dataset. We show the estimates of both observed and counterfactual
  mean outcomes along with the estimates of the ATT and ATC.}
  \label{tab:bmc_att_atc}
\end{table}

The results are shown in Table~\ref{tab:bmc_att_atc}. Our outcome
variable of interest is the \emph{change} in HbA1c level, therefore
the estimated $\E[Y(1)]$ and $\E[Y(0)]$ represent our estimates of the
change in HbA1c for the treated groups and untreated groups
specifically, while ATT and ATC records the estimated change in HbA1c
that results from the treatment.  We find that the estimates based on
Causal Forests and the doubly-robust estimator have negative estimates
of both ATT and ATC.  This suggests that being prescribed Metformin
lowers the level of blood glucose, benefiting patients.

However, the negative ATC estimate is primarily due to a large
negative estimate of the counterfactual mean outcome, i.e.,
$\E[Y(1) \mid T=0]$.  Although the lack of experimental data makes it
difficult to assess the accuracy of these estimates, it is highly
unlikely that any single individual can produce a change of $1.5$
points or greater in HbA1c in a single period.  Indeed, under the
standard settings, a reduction of HbA1c by $0.5$ points or greater is
already considered clinically significant
\citep{spaulonci2013randomized}. Thus, the absolute outcome estimates
of Causal Forests and the doubly robust estimator appear to be
overstating the efficacy of MET.

In contrast, the estimates based on the distributional sensitivity
model are much more realistic.  The ATT estimate shows an effect of
moderate size while the ATC estimate is slightly positive.  The
results therefore suggest that the effectiveness of MET is far from
conclusive.

\subsection{STAR Dataset}

As the third and final application, we analyze the STAR dataset using
the distributional sensitivity model. The STAR project is a randomized
experiment stratified on the school level that randomly assigned over
7,000 students across 79 schools to three different groups: small
class, regular class without aid, and regular class with a full-time
teacher's aid \citep{most:95}. The experiment began when students
entered kindergarden and continued through third grade, with
assessments at the end of each grade.

We focus on the comparison of two groups --- small class and regular
class without aid.  Our binary treatment assignment variable is equal
to 1 if a student is assigned to a small class and is 0 otherwise.
For the outcome variable, we consider the reading score at the end of
kindergarten. There are a total of ten pre-treatment covariates,
including four demographic characteristics of students (gender, race,
birth month, birth year) and six school characteristics (urban/rural,
enrollment size, grade range, number of students on free lunch, number
of students on school buses, and percentage of white students).

We construct a non-experimental control group from the STAR by broadly
following the strategy of \cite{wilde2007close}. First, we subset the
data to select 28 schools that had over 50 participating students
during kindergarten. Second, to create a non-experimental control
group for each school, we randomly permute at the school level for the
students in the control group to simulate a setting where the control
group does not come from experimental data (in this case coming from
experimental data from \emph{another} school). Third, for each school,
we apply the distributional sensitivity model to robustly estimate the
school-level ATT.  Finally, we compute the stratified ATT estimate by
weighting the school-level ATT estimates across all schools by the
number of student participants.

\begin{table}[t]
  \centering
  \begin{tabular}{|..|rrr|}
    \hline \multicolumn{1}|{r}{\quad \ $\bm{\Gamma}$} & \multicolumn{1}{c|}{$\bm{\delta}$} & \multicolumn{1}{c}{\textbf{Estimate}} & \multicolumn{1}{c}{\textbf{s.e.}} & \multicolumn{1}{c|}{\textbf{Bias}}  \\\hline
    1.1 & 0.3 & 5.67 & 1.38 & $-$2.01\\
    1.25 & 0.3 & 4.27 & 1.42 & $-$3.41\\
    1.4 & 0.3 & 0.76 & 1.46 & $-$6.92\\
    1.6 & 0.3 & $-$2.33 & 1.52 & $-$10.01\\\hline       
  \end{tabular}
  \caption{Results of the distributional sensitivity model on the STAR
    dataset under different robustness parameters.}
  \label{tab:star_results}
\end{table}

The results are shown in Table~\ref{tab:star_results}.  Standard
errors are calculated using the same method as in
Section~\ref{subsec:nsw}.  We find that when with $\Gamma=1.1$, the
estimate is once again not statistically significantly different from
the experimental benchmark. As $\Gamma$ grows, however, the estimated
ATT quickly becomes negative. This is due to the fact that each school
has a small number of samples (50--100 students), and thus a larger
$\Gamma$ results in an estimate based on the unidentifiable
distributions with just a few samples.  This suggests that with a
small sample size, the distributional sensitivity model may yield an
overly conservative estimate.

\section{Conclusion}  
  
In this work, we proposed a new robust causal inference framework with
distributionally robust optimization. We showed how the commonly used
marginal sensitivity model is a special case of our framework, and
proposed a new model, the distributional sensitivity model, that
overcomes its limitation and has a wide applicability. We illustrate,
on both synthetic and empirical data, that the distributional
sensitivity model is able to generate estimates that are only
moderately conservatively biased and offers robust guarantees.

\bibliography{imai, my, obsinf}
\clearpage
\appendix

\setcounter{table}{0}
\renewcommand{\thetable}{A\arabic{table}}
\setcounter{figure}{0}
\renewcommand{\thefigure}{A\arabic{figure}}
\setcounter{equation}{0}
\renewcommand{\theequation}{A\arabic{equation}}
\section{Supplementary Appendix}

\subsection{Proof of Proposition~\ref{prop:marginalsensitivity}}
\label{app:marginalsensitivity}
  
\begin{proof}
  By algebraic manipulation, the marginal sensitivity model is
  equivalent to:
\begin{gather*}
	\min_{\pi(\bU_i)} \frac{1}{n_1} \sum_{i: T_i=1} Y_i - \frac{p_0}{n_0}\sum_{i: T_i=0} Y_i \frac{\pi(\bU_i)}{1-\pi(\bU_i)}  \\
s.t. \quad \frac{p_1}{\Gamma p_0} \leq \frac{\pi(\bU_i)}{1-\pi(\bU_i)}\leq \frac{\Gamma p_1}{p_0}  \; \; \forall i \\
	\sum_{i: T_i=0} \frac{\pi(\bU_i)/(1-\pi(\bU_i))}{p_1/p_0} = n_0,\qquad
\sum_{i: T_i=1} \frac{p_1/p_0}{\pi(\bU_i)/(1-\pi(\bU_i))} = n_1
\end{gather*}
where $n_t=np_t$ for $t=0,1$.  
Define $\lambda_i = p_0 \pi(\bU_i)/\{p_1(1-\pi(\bU_i))\}$ and we can  
rewrite the above formulation, after some rearrangement, as: 
\begin{gather}
	\min_{\lambda_i} \frac{1}{n_1} \sum_{i: T_i=1} Y_i - \sum_{i: T_i=0} \frac{\lambda_i}{n_0} Y_i   \label{eq:inference_msm} \\
	s.t. \quad \frac{1}{\Gamma} \leq \lambda_i \leq \Gamma  \; \; \forall i \label{eq:inference_msm_cons1}	\\	\sum_{i: T_i=0} \frac{\lambda_i}{n_0} = 1,\qquad 
	\sum_{i: T_i=1} \frac{1}{n_1\lambda_i} = 1\label{eq:inference_msm_cons2}	.
\end{gather}
Now, consider the empirical version of the distributionally robust
causal inference problem for the ATT under our framework shown in
Equation~\eqref{eq:att}:
\begin{equation}
	\min_{f_{01},f_{10} \in S}  \frac{1}{n_1}\sum_{i:T_i=1} Y_i -\int y f_{01}(y) \d y \label{eq:emp_att}
\end{equation}
Therefore, we find that the problem in
Equations~\eqref{eq:inference_msm}--\eqref{eq:inference_msm_cons2}
correspond exactly to the distributionally robust causal inference
problem under our framework with the following ambiguity sets:
\begin{align*}
    &S^{MS}_{01}(\Gamma)=\left\{ f_{01}(y) \mid F_{01}(y)=\widehat{F}_0(y; \mathbf{w}),  \;\;\sum_{i:T_i=0} w_i = 1,\;\;\frac{1}{\Gamma} \leq n_0w_i \leq \Gamma \;\;\forall i,j \right\}
\end{align*}
\end{proof}

\subsection{Proof of Proposition \ref{prop:distributional_bound}}

Denote $\mu_{01}=\E[Y_i(1)-Y_i(0) \mid T_i = 1]$. Then, for any $\epsilon>0$, we have:
\begin{align*}
  & \ \min_c KS(F_{01}(y), F_{11}(y+c)) \\
  \leq  & \  KS(F_{01}(y),
  F_{11}(y+\mu_{01})) \\  = & \ \max_c |\P(Y_i(0) \leq c \mid T_i = 1)
                              - \P(Y_i(1) \leq c + \mu_{01} \mid T_i =
                              1)|. \\
	\leq & \ \max_c |\P(Y_i(0) \leq c \mid T_i = 1) - \P(Y_i(0) \leq c - \epsilon \wedge Y_i(1)-Y_i(0) \leq \epsilon + \mu_{01} \mid T_i = 1)|
	\\ \leq & \ \max_c |\P(Y_i(0) \leq c \mid T_i = 1) - \P(Y_i(0) \leq c - \epsilon \mid T_i = 1) - \P(Y_i(1)-Y_i(0) \leq \epsilon + \mu_{01} \mid T_i = 1) + 1|
	\\ = & \ \max_c \{\P(Y_i(0) \leq c \mid T_i = 1) - \P(Y_i(0) \leq c
  - \epsilon \mid T_i = 1)\} + 1-\P(Y_i(1)-Y_i(0) \leq \epsilon +
  \mu_{01} \mid T_i = 1) \\
	\leq & \ k\epsilon + (1-\P(Y_i(1)-Y_i(0) \leq \epsilon + \mu_{01}
   \mid T_i = 1)) \\
	\leq & \ k\epsilon + \frac{\sigma_{01}^2}{\epsilon^2}
\end{align*}
where the last two inequalities follow from the $k$-Lipschitz
condition and Chebyshev's inequality.  Now, since this is true for all
$\epsilon>0$, we minimize this expression over $\epsilon$. The minimum
is reached at $\epsilon=(\frac{2\sigma^2}{k})^{1/3}$, and thus we
have:
\[\min_c KS(F_{01}(y), F_{11}(y+c) \leq
  3\left(\frac{k\sigma_{01}}{2}\right)^{2/3}\]

\subsection{Proof of Theorem \ref{thm:metric_duality}}
\label{app:metric_duality}

We begin by defining the following two ambiguity sets:
\begin{align*}
    S^{DSa}_{01}(\Gamma, \delta)&=\left\{ f_{01}(y) \mid F_{01}(y)=\widehat{F}_0(y; \mathbf{w}),  \;\;\sum_{i:T_i=0} w_i = 1,\right.\\
    &\left.\;\; \qquad \qquad 0\leq w_i \leq \frac{\Gamma}{n_0} \;\;\forall i,  \;\; \min_c KS(F_{01}(y),F_{11}(y+c)) \leq \delta \right\},
\end{align*}
and
\begin{align*}
    S^{DSb}_{01}(\Gamma, \delta)&=\left\{ f_{01}(y) \mid F_{01}(y)=\widehat{F}_1(y; \mathbf{w}),  \;\;\sum_{i:T_i=0} w_i = 1, \;\;w_i\geq 0,\right.\\
    &\left.\;\;   \;\; \qquad \qquad d_0(F_{01}(y),F_{00}(y))\leq \frac{\Gamma - 1}{n_0}, d_1(F_{01}(y),F_{11}(y)) \leq \delta \right\}.
\end{align*}
Now, suppose $f_{01}(y)\in S^{DSa}_{01}(\Gamma, \delta)$. Then, we know
that
\[F_{01}(y)=\widehat{F}_t(y; \mathbf{w})=\sum_{i:T_i=0}
  w_iF_{Y_i}(y),\] where $\sum_{i:T_i=0} w_i = 1$,
$F_{Y_i}(y)=\bm{1}\{y\geq Y_i\}$, and
$0\leq w_i \leq \frac{\Gamma}{n_0}$ holds. 
Let us first consider the case where $\Gamma \geq 2$. Then we have, by definition of $d_0$:
\begin{align*}
    d_0(F_{01}(y),F_{00}(y))& = \max_{i} \left|w_i-\frac{1}{n_0}\right|\\& \leq \max\left\{\frac{\Gamma-1}{n_0},\frac{1}{n_0}\right\}\\& \leq \frac{\Gamma-1}{n_0} .
\end{align*}
where we used the fact that $F_{00}(y)=\sum_{i:T_i=0}
F_{Y_i}(y)/n_0$. Therefore, we have
$f_{01}(y)\in S^{DSb}_{01}(\Gamma, \delta)$.  For the case where
$\Gamma<2$, we have:
\begin{align*}
    d_0(F_{01}(y),F_{00}(y))& = \max_{i} \{w_i-\frac{1}{n_0},0\}\\& \leq \frac{\Gamma-1}{n_0}.
\end{align*}
Together, we have shown,
$f_{01}(y)\in S^{DSa}_{01}(\Gamma, \delta) \Rightarrow f_{01}(y)\in
S^{DSb}_{01}(\Gamma, \delta)$.

Next, we prove the reverse direction. Let
$f_{01}(y)\in S^{DSb}_{01}(\Gamma, \delta)$. Again, let us first
consider the case where $\Gamma \geq 2$. By definition of
$S^{DSb}_{01}(\Gamma, \delta)$, we have:
\[d_0(F_{01}(y),F_{00}(y)) = \max_{i}
  \left|w_i-\frac{1}{n_0}\right|\leq \frac{\Gamma-1}{n_0}\] This
implies $\max_i w_i \leq \Gamma/n_0$.  Thus, we have that
$f_{01}(y)\in S^{DSa}_{01}(\Gamma, \delta)$. For the case where
$\Gamma<2$, by definition of $S^{DSb}_{01}(\Gamma, \delta)$, we have:
\[d_0(F_{01}(y),F_{00}(y)) = \max_{i}
  \max\left\{w_i-\frac{1}{n_0},0\right\}\leq \frac{\Gamma-1}{n_0},\]
which similarly implies $\max_i w_i \leq \Gamma/n_0$.  Therefore, we
have shown
$f_{01}(y)\in S^{DSb}_{01}(\Gamma, \delta) \Rightarrow f_{01}(y)\in
S^{DSa}_{01}(\Gamma, \delta)$, which, together with the above result,
implies that
$S^{DSb}_{01}(\Gamma, \delta) = S^{DSa}_{01}(\Gamma, \delta)$.

We now prove that $d_0$ is a metric on the space of piecewise constant
CDFs when $\Gamma \geq 2$ and a quasimetric when $\Gamma <2$. A metric
$d$ must satisfy three properties:
\begin{enumerate}
    \item $d(x,y)=0 \Leftrightarrow x=y$
    \item $d(x,y)=d(y,x)$
    \item $d(x,y)\leq d(x,z)+d(z,y)$
\end{enumerate}
A quasimetric $d'$ needs to only satisfy properties 1 and 3. 

We first consider the $\Gamma\geq 2$ case. By symmetry of the absolute
value function, $d_0$ satisfies property 2. For property 1, if $F=G$,
it is clear that $d(F,G)=0$. If $d(F,G)=0$, then it implies:
\[\max_y \left|\lim_{\epsilon \to 0^-} (F(y)-G(y))-(F(y+\epsilon)- G(y+\epsilon))\right|=0\]
which leads to:
\begin{equation}
    \lim_{\epsilon \to 0^-} (F(y)-G(y))-(F(y+\epsilon)- G(y+\epsilon))=0 \quad \forall y \label{eq:limit1}
\end{equation}
Note that $F, G$ are piecewise constant CDFs. Therefore, $F$ can be completely characterized by its $n_F$ jump points:
\[F(y)= \sum_{i=1}^{n_F}w_i \bone\{y>y_i\},\] where
$\sum_{i=1}^{n_F}w_i =1$. Then, we apply Equation~\eqref{eq:limit1} at
$y=y_i$ for all $i \in \{1,\cdots,n_F\}$, which yields:
\[\lim_{\epsilon \to 0^-} G(y)-G(y+\epsilon)=w_i\]
That is, $G$ also has a jump of magnitude $w_i$ at $y_i$ for all
$i$. Since $\sum_{i=1}^{n_F} w_i=1$, this means that $G$ cannot have
any additional discontinuity points, and thus we have:
\[G(y)= \sum_{i=1}^{n_F}w_i \bone\{y>y_i\}=F(y).\]

For property 3, we have:
\begin{align*}
    d_0(F,G)+d_0(G,H)&=\max_{y_1} \left|\lim_{\epsilon \to 0^-} (F(y_1)-G(y_1))-(F(y_1+\epsilon)- G(y_1+\epsilon))\right|\\&+\max_{y_2} \left|\lim_{\epsilon \to 0^-} (G(y_2)-H(y_2))-(G(y_2+\epsilon)- H(y_2+\epsilon))\right|
    \\& \geq \max_{y_1} \left|\lim_{\epsilon \to 0^-} (F(y_1)-G(y_1))-(F(y_1+\epsilon)- G(y_1+\epsilon))\right|\\&+ \left|\lim_{\epsilon \to 0^-} (G(y_1)-H(y_1))-(G(y_1+\epsilon)- H(y_1+\epsilon))\right|\\& \geq \max_{y_1} \left|\lim_{\epsilon \to 0^-} (F(y_1)-G(y_1))-(F(y_1+\epsilon)- G(y_1+\epsilon))\right.\\&+\biggl. (G(y_1)-H(y_1))-(G(y_1+\epsilon)- H(y_1+\epsilon))\biggr|\\&= \max_{y_1} \left|\lim_{\epsilon \to 0^-} (F(y_1)-H(y_1))-(F(y_1+\epsilon)- H(y_1+\epsilon))\right|\\&=d_0(F,H),
\end{align*}
as required. Therefore $d_0$ is a metric for $\Gamma \geq 2$.

We next prove the case where $\Gamma<2$. Again, it is clear that if
$F=G$, then we have $d_0(F,G)=0$. If $d_0(F,G)=0$, then we have:
\[\max_y \max\left\{\lim_{\epsilon \to 0^-} (F(y)-G(y))-(F(y+\epsilon)- G(y+\epsilon)),0\right\}=0\]
This implies that:
\begin{equation}
    \lim_{\epsilon \to 0^-} (F(y)-G(y))-(F(y+\epsilon)- G(y+\epsilon))\leq 0 \quad \forall y \label{eq:limit2}
\end{equation}
\[\lim_{\epsilon \to 0^-} (F(y)-G(y))-(F(y+\epsilon)- G(y+\epsilon))\leq 0 \quad \forall y\]
Let us consider the characterization of $F$ by its $n_F$ jump points:
\[F(y)= \sum_{i=1}^{n_F}w_i \bone\{y>y_i\},\] If we apply
Equation~\eqref{eq:limit2} to the $y_i$ for all
$i \in \{1,\cdots, n_F\}$, and then sum up the equations, we have:
\[\lim_{\epsilon \to 0^-} \sum_{i=1}^{n_F} G(y_i)-G(y_i+\epsilon)\geq 1\]
However, since $G$ is a CDF, we have:
\[\lim_{\epsilon \to 0^-} \sum_{i=1}^{n_F} G(y_i)-G(y_i+\epsilon)\leq G(\max_i y_i)-G(\min_i y_i) \leq  1\]
Therefore, we have:
\[\lim_{\epsilon \to 0^-} \sum_{i=1}^{n_F} G(y_i)-G(y_i+\epsilon)= 1\]
Now assume that we have
$\lim_{\epsilon \to 0^-} G(y_i)-G(y_i+\epsilon)<w_i$ for some
$i$. Then, by Equation~\eqref{eq:limit2} at $y=y_i$, we have:
\begin{equation*}
    \lim_{\epsilon \to 0^-} (F(y_i)-G(y_i))-(F(y_i+\epsilon)- G(y_i+\epsilon))=w_i-(G(y_i)-G(y_i+\epsilon))> 0,
\end{equation*}
which leads to a contradiction. Therefore, we have that
$\lim_{\epsilon \to 0^-} G(y_i)-G(y_i+\epsilon)\geq w_i$ for all
$i$. But then since $\sum_{i=1}^{n_F} w_i =1$, $G$ thus cannot have
any additional jump points and therefore:
\[G(y)= \sum_{i=1}^{n_F}w_i \bone\{y>y_i\}=F(y),\]
as required.

For property 3, we have:
\begin{align*}
    d_0(F,G)+d_0(G,H)&=\max_{y_1} \max\left\{\lim_{\epsilon \to 0^-} (F(y_1)-G(y_1))-(F(y_1+\epsilon)- G(y_1+\epsilon)),0\right\}\\&+\max_{y_2} \max\left\{\lim_{\epsilon \to 0^-} (G(y_2)-H(y_2))-(G(y_2+\epsilon)- H(y_2+\epsilon)),0\right\}
    \\& \geq \max_{y_1} \max\left\{\lim_{\epsilon \to 0^-} (F(y_1)-G(y_1))-(F(y_1+\epsilon)- G(y_1+\epsilon)),0\right\}\\&+ \max\left\{\lim_{\epsilon \to 0^-} (G(y_1)-H(y_1))-(G(y_1+\epsilon)- H(y_1+\epsilon)),0\right\}\\& \geq \max_{y_1}\max\left\{ \lim_{\epsilon \to 0^-} (F(y_1)-H(y_1))-(F(y_1+\epsilon)- H(y_1+\epsilon)),0\right\}\\&=d_0(F,H).
\end{align*} 
Thus, $d_0$ is a quasimetric when $\Gamma <2$.

\newpage
\subsection{Additional Synthetic Experiments on Varying $\delta$}
\label{app:synthetic_delta}

\begin{table}[h!]
	\centering\setlength{\tabcolsep}{2.5pt}
	\resizebox{\textwidth}{!}{
		\begin{tabular}{l|..|..|..}
			\hline
			&  \multicolumn{2}{c|}{$\bm{n=100}$} &  \multicolumn{2}{c|}{$\bm{n=200}$}& \multicolumn{2}{c}{$\bm{n=500}$}\\\hline
			& \multicolumn{1}{c}{bias} &
			\multicolumn{1}{c|}{s.d.}
			& \multicolumn{1}{c}{bias} &
			\multicolumn{1}{c|}{s.d.}
			& \multicolumn{1}{c}{bias} &
			\multicolumn{1}{c}{s.d.}  \\ \hline
			\multicolumn{1}{l|}{\textbf{Scenario 1}}  & & & & & &\\
			\multicolumn{1}{l|}{\textit{Distributional Sensitivity}}  & & & &&&\\
			\quad $\Gamma = 2$  &-1.810 & 0.498  & -1.888 & 0.335   &  -1.913  & 0.213   \\
			\quad $\Gamma = 3$ & -2.245  & 0.543  & -2.385& 0.384    &  -2.455 &  0.234 \\
			\quad $\Gamma = 5$   & -2.729 & 0.649  & -2.900 & 0.462  &  -2.900  &  0.275   \\
			\multicolumn{1}{l|}{\textit{Marginal Sensitivity}}  & & & &&&\\
			\quad $\Gamma = 2$  & -1.938 & 0.465  & -1.954 & 0.316  &  -1.944  & 0.207   \\
			\quad $\Gamma = 3$ & -2.506  & 0.470  & -2.506 & 0.348    &  -2.550 &  0.218   \\
			\quad $\Gamma = 5$   & -3.164 & 0.529  & -3.175 & 0.391  &  -3.192  &  0.241   \\\hline
			\multicolumn{1}{l|}{\textbf{Scenario 2}}  & & & & & &\\
			\multicolumn{1}{l|}{\textit{Distributional Sensitivity}}  & & & &&&\\
			\quad $\Gamma = 2$  & -1.190 & 0.489  & -1.263& 0.336   &  -1.302  & 0.213  \\
			\quad $\Gamma = 3$ & -1.653  & 0.543  & -1.797 & 0.369    &  -1.882&  0.234   \\
			\quad $\Gamma = 5$   & -2.094  & 0.661  & -2.296 & 0.465 &  -2.412  &  0.274   \\
			\multicolumn{1}{l|}{\textit{Marginal Sensitivity}}  & & & &&&\\
			\quad $\Gamma = 2$  & -1.321 & 0.455  & -1.330 & 0.321  &  -1.333  & 0.208  \\
			\quad $\Gamma = 3$ & -1.916  & 0.474  & -1.953 & 0.339    & -1.973 &  0.216   \\
			\quad $\Gamma = 5$   & -2.518 & 0.546  & -2.577 & 0.394  &  -2.594  &  0.233   \\\hline
			\multicolumn{1}{l|}{\textbf{Scenario 3}}  & & & & & &\\
			\multicolumn{1}{l|}{\textit{Distributional Sensitivity}}  & & & &&&\\
			\quad $\Gamma = 2$  & -1.667 & 0.553  & -1.788 & 0.370  &  -1.831  & 0.226   \\
			\quad $\Gamma = 3$ & -2.113  & 0.632  & -2.260 & 0.441    & -2.369 &  0.263   \\
			\quad $\Gamma = 5$   & -2.457  & 0.752  & -2.745 & 0.519 & -2.882  &  0.326   \\
			\multicolumn{1}{l|}{\textit{Marginal Sensitivity}}  & & & &&&\\
			\quad $\Gamma = 2$  & -1.847 & 0.513  & -1.870 & 0.345  &  -1.865  & 0.218   \\
			\quad $\Gamma = 3$ & -2.466 & 0.549  & -2.455 & 0.392    & -2.475 &  0.244   \\
			\quad $\Gamma = 5$   & -3.047 & 0.598  & -3.090 & 0.435  &  -3.092  &  0.271   \\\hline\hline
		\end{tabular}
	}
	\caption{A Simulation Study of Distributional Sensitivity (with $\delta = 0.15$) and
		Marginal Sensitivity Models.  The table presents the estimated
		bias and standard deviation of the ATT estimators for both models under difference scenarios, robustness ($\Gamma$) and number of samples $n$.} \label{tb:synthetic_ms_compare_add}
\end{table}
\end{document}